\documentclass[runningheads]{llncs}
\usepackage{booktabs} 

\usepackage{listings}
\usepackage{graphicx}
\PassOptionsToPackage{hyphens}{url}
\usepackage[unicode]{hyperref}
\hypersetup{
    pdfauthor={Klaus Schmid; Sascha El-Sharkawy; Christian Kröher},
    pdftitle={Improving Software Engineering Research through Experimentation Workbenches},
    pdfsubject={From Software Engineering to Formal Methods and Tools, and Back --- Essays Dedicated to Stefania Gnesi on the Occasion of Her 65th Birthday (LNCS, volume 11865)},
    pdfkeywords={Experimentation Workbench; Empirical Software Engineering, Static Analysis, Software Product Line Analysis},
    pdfpagemode={UseOutlines},
    pdfproducer={LaTeX},
		bookmarksopenlevel=2,
		colorlinks=true,
    linkcolor=black,            
    citecolor=black,            
    filecolor=black,            
    urlcolor=black              
}
\usepackage{url}
\expandafter\def\expandafter\UrlBreaks\expandafter{\UrlBreaks
  \do\a\do\b\do\c\do\d\do\e\do\f\do\g\do\h\do\i\do\j%
  \do\k\do\l\do\m\do\n\do\o\do\p\do\q\do\r\do\s\do\t%
  \do\u\do\v\do\w\do\x\do\y\do\z\do\A\do\B\do\C\do\D%
  \do\E\do\F\do\G\do\H\do\I\do\J\do\K\do\L\do\M\do\N%
  \do\O\do\P\do\Q\do\R\do\S\do\T\do\U\do\V\do\W\do\X%
  \do\Y\do\Z}
\usepackage[utf8]{inputenc}

\usepackage{xcolor} 

\definecolor{pink}{rgb}{0.9,0,0.9}
\definecolor{gray}{rgb}{0.95,0.95,0.85}
\definecolor{ass}{rgb}{0.95,0.25,0.85}
\definecolor{mauve}{rgb}{0.1,0.7,0.2}
\definecolor{blue}{rgb}{0,0,0.95}
\definecolor{green}{rgb}{0,0.95,0}
\definecolor{RED}{rgb}{0.95,0,0}
\definecolor{orange}{rgb}{1,0.5,0}

\definecolor{cEclipseGreen}{rgb}{0.25,0.5,0.35}
\definecolor{cEclipseBlue}{rgb}{0.25,0.35,0.75}

\usepackage{listings}
\lstdefinelanguage{properties} {
    tabsize=1,
    commentstyle=\color{blue}, 
    morecomment=[l]{\	},
    morecomment=[l]{)},    
    keywordstyle=[2]{\color{black}},
    morekeywords=[2]{=}
}

\usepackage{enumitem}
\newcommand{\RQLayerOne}[1]{\textit{R#1}}
\setlist[itemize]{leftmargin=6ex}
\setlist[enumerate]{leftmargin=6ex}

\usepackage[absolute,overlay]{textpos}
\usepackage{mdframed}

\newcommand{\AuthorsVersion}{
\noindent \begin{mdframed}\scriptsize
		\textit{Improving Software Engineering Research through Experimentation Workbenches}\\
		This is the author’s version of the work. It is posted here for your personal use.\\Not for redistribution. The definitive Version of Record was published in\\
		\textit{From Software Engineering to Formal Methods and Tools, and Back --- Essays Dedicated to Stefania Gnesi on the Occasion of Her 65th Birthday (LNCS, volume 11865), 2019,} \url{http://dx.doi.org/10.1007/978-3-030-30985-5_6}
	\end{mdframed}
}

\begin{document}
\title{Improving Software Engineering Research through Experimentation Workbenches}
\titlerunning{Improving SE Research through Experimentation Workbenches}

\author{Klaus Schmid \and Sascha El-Sharkawy \and Christian Kröher}

\institute{University of Hildesheim, Institute of Computer Science, Hildesheim, Germany
	\email{\{schmid,elscha,kroeher\}@sse.uni-hildesheim.de}\\
	\url{https://sse.uni-hildesheim.de/en/}
}
\authorrunning{K. Schmid et al.}


\maketitle

\begin{abstract}
Experimentation with software prototypes plays a fundamental role in software engineering research. In contrast to many other scientific disciplines, however, explicit support for this key activity in software engineering is relatively small. While some approaches to improve this situation have been proposed by the software engineering community, experiments are still very difficult and sometimes impossible to replicate.

In this paper, we propose the concept of an \textit{experimentation workbench} as a means of explicit support for experimentation in software engineering research. In particular, we discuss core requirements that an experimentation workbench should satisfy in order to qualify as such and to offer a real benefit for researchers. Beyond their core benefits for experimentation, we stipulate that experimentation workbenches will also have benefits in regard to reproducibility and repeatability of software engineering research. Further, we illustrate this concept with a scenario and a case study, and describe relevant challenges as well as our experience with experimentation workbenches.

\end{abstract}

\begin{textblock*}{\textwidth}(4.75cm,24cm) 
   \AuthorsVersion
\end{textblock*}
\section{Introduction}
A significant part of software engineering is experimental in nature. This holds both for method-oriented research, which typically requires humans-in-the-loop, as well as  more implementation-oriented research (related to program analysis, verification, software generation, etc.), which is the focus of this contribution. 

The challenges to experimental research in software engineering are very similar to these in other experimental disciplines, like physics or psychology. Those include replicability of research results, efficient support for the experimental process, like conducting variations, or enabling others to reuse the scientific results. In some disciplines these issues have gained wide-spread attention, like in psychology due to the reproducibility crisis~\cite{Baker15}. In large-scale physics, like the Large Hadron Collider (LHC), creating documentation solutions and supporting many variations of experiments is considered well before any experiments are actually built, i.e., creating the experiments are major systematic engineering activities in their own right. This inspired us to compare this situation with software engineering research, in particular experimental research based on software tools. 

In software engineering, deficiencies in the systematic support of the research process are increasingly recognized as an issue. In our own experience (and that of others), even if the relevant software is provided, e.g., as open-source, it is very difficult and sometimes impossible to replicate the experiments as they may rely on (unavailable) third party tools or  undocumented execution details. Thus, the replication of a single evaluation may require several days or weeks of work only for reverse engineering missing information or assets. This has also influenced organizations, like the Association for Computing Machinery (ACM), to address this need and provide guidelines to improve the situation, e.g., with assessing publications~\cite{badging}. As part of these guidelines, ACM defines a terminology that distinguishes repeatability, replicability, and reproducibility. In this paper, we will follow this terminology and, hence, use these terms as follows:
\begin{itemize}
	\item \textit{Repeatability} means that researchers receive the same results with their own experimental setup on multiple trials.
	\item \textit{Replicability} means that a different person receives the same results with the same experimental setup as reported by a researcher on multiple trials.
	\item \textit{Reproducibility} means that a different person receives the same results as reported by a researcher with their own experimental setup on multiple trials.
\end{itemize}

A typical way to improve repeatability, replicability, and reproducibility is the publication of all artifacts relevant to an experiment. For instance, conferences increasingly provide the possibility to back up publications with artifacts and assess their quality~\cite{badging}. Other  measures include the use of docker or virtual machines to improve replicability~\cite{Boettiger15}.
However, these approaches are typically applied after the fact, i.e., after the experiments are finished, as opposed to practices in  established experimental disciplines. This post-mortem approach may lead to  threats to validity as it leads to the risk of missing important details in the documentation artifacts. These solutions do also not address other issues in the scientific process, like exploration of experimental variation. 

Here, driven from our own experiences in conducting technical research experiments, we propose the concept of an \textit{experimentation workbench} for software engineering to remedy this situation and make the scientific workflow and its requirements a  central aspect in the tools we build. A key motivation for our proposal is the question:

\begin{quotation}
\noindent\textit{``How would a support environment for software engineering research look like, if we would specifically engineer one?''}
\end{quotation}

Today, we are used to \textit{development workbenches} like Eclipse~\cite{Eclipse}, but while they are heavily used in research, they (only) aim at supporting the software development process in general. They do not address any specific research-oriented requirements.
Other uses of the term workbench include artifacts, like \textit{language workbenches}~\cite{DyerNguyenRajan+13}. 
Again, this term is    more directed towards (language)  development, not so much towards research. 
We choose the term \textit{experimentation workbench} in analogy to these uses of the term. 
The term experimentation workbench is also not  completely new. It has already been used in networking~\cite{EideStollerLepreau07}, however, with slightly different semantics, namely to denote a specific form of simulation environment.

An experimentation workbench, as we envision it, is not only about replicability, but about supporting the scientific process at large (e.g., rapid variation, reuse in new research), as we will discuss in the following sections. Thus, among other things, it should also support general reproducibility.
This would move software engineering more in line with other experimental sciences. The requirements we put forward for defining the concept of experimentation workbenches are our main contribution. We believe thinking in these terms from the beginning and supporting the scientific process with such environments can be a major contribution to our community. In summary, our contributions  are:
\begin{itemize}
	\item The definition of the concept of an experimentation workbench  along with a description of its defining requirements.
	\item An illustrative scenario highlighting the benefits of experimentation workbenches.
	\item An example implementation (KernelHaven).
	\item A discussion of challenges for creating experimentation workbenches.
	\item A report of our experiences with realizing and using experimentation workbenches in our research on product line analysis.
\end{itemize}

Below, we will further refine the concept of experimentation workbenches in a scenario (Section~\ref{sec:scenario}), before we define the fundamental requirements in Section~\ref{sec:requirements}. We illustrate the defined concept based on KernelHaven in Section~\ref{sec:implementation}, discuss major challenges to realizing experimentation workbenches in Section~\ref{sec:discussion}, and provide our experiences in Section~\ref{sec:experiences}. Finally, we conclude in Section~\ref{sec:conclusion}.

\section{Usage Scenario}\label{sec:scenario} 

\begin{figure}[t] 
	\centering
		\includegraphics[width=\columnwidth,trim={0cm 11,3cm 11cm 0cm},clip]{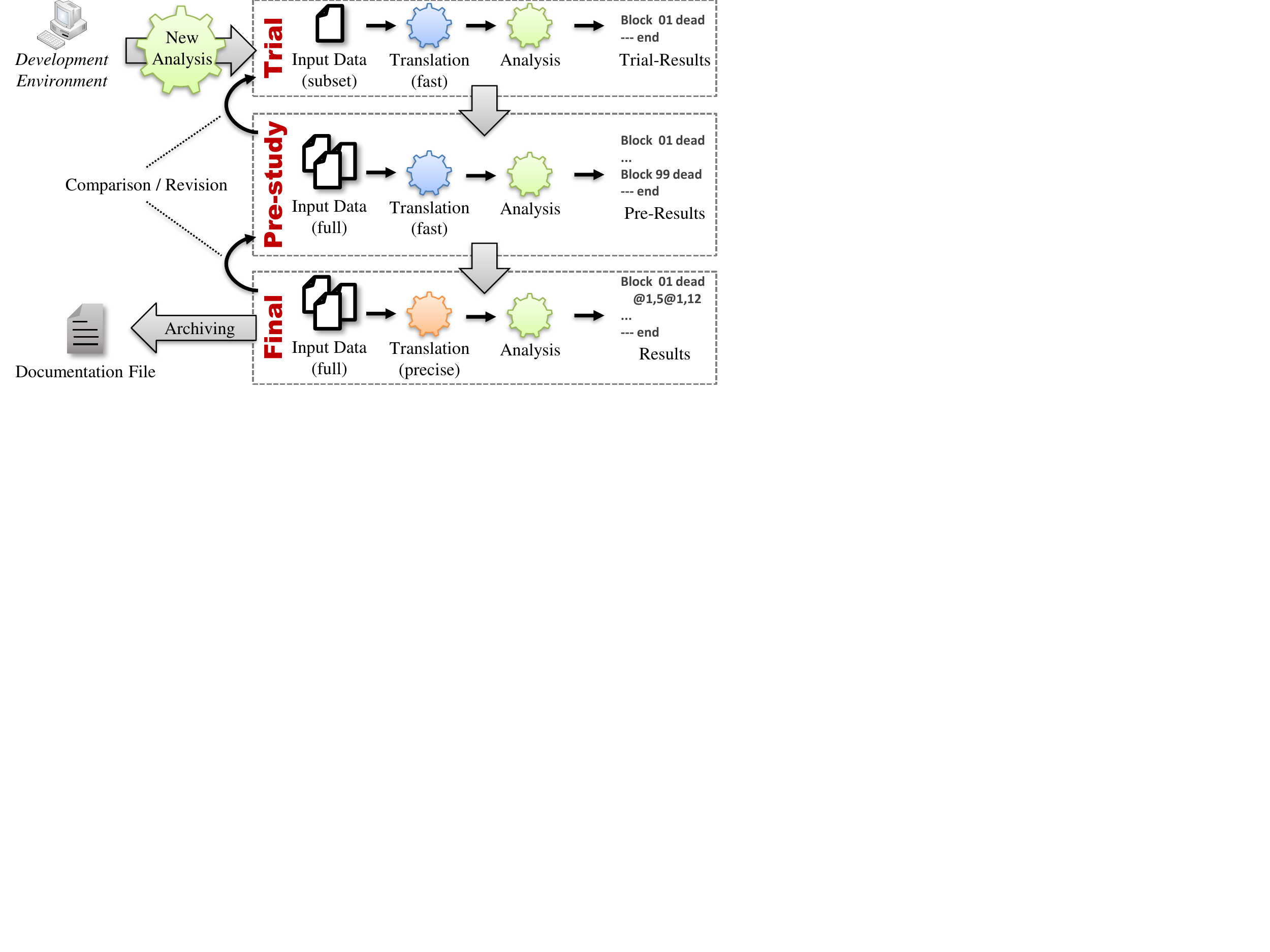}
	\caption{Example workflow using an experimentation workbench.}\label{fig:Scenario}
\end{figure}
In this section, we describe a scenario to clarify our expectations on how experimentation workbenches support the  experimentation workflow. We assume experimentation workbenches to be constructed for a specific research domain. For our scenario we use the domain of static product line analysis as a reference, which is a rather active field of research~\cite{ThuemApelKaestner+14}. It aims at questions like detecting code that can never be part of a product, as there is no product configuration that would allow this, or detecting type inconsistencies that only arise for specific code configurations. All these analyses have a certain structure: the different inputs like a variability model, source code, etc.\ must be analyzed and transformed into  appropriate formats for integration and  analysis. We choose this domain to match it to the example experimentation workbench discussed in Section~\ref{sec:implementation}. Figure~\ref{fig:Scenario} shows an illustration of an example workflow in this domain as supported by an experimentation workbench. We discuss this workflow below by means of a scenario.

\textit{Preparation}. Stefania wants to test her new analysis approach. She implemented it as a plugin for an experimentation workbench. This analysis works on an abstract representation of a product line and requires inputs from the variability model, the variability-enhanced build model, and C-code files. She uses Linux as a case study, which is often used  in research, but huge. For translating the source code into an appropriate format for her analysis, two techniques are available: a fast, but not so precise one, and one precise, but rather slow.

\textit{Trial}. First, Stefania  wants to perform a trial with a  small subset of the data. Thus, she  defines the case study subset, the fast translation technique, and her analysis by configuration of the experimentation workbench. This is possible as data format standardization (along with necessary translations) and other services are offered by the experimentation workbench. The workbench also addresses parallelization and other technical issues regarding resource utilization allowing her to focus only on the realization of her analysis. 
In particular, no coding is required (except for implementing her analysis). This run gives the expected results after a few minutes.

\textit{Pre-study}. Stefania changes the   configuration to include all input data for her pre-study. She starts the analysis, which finishes already in a few hours. The results are again positive, but some files have not been correctly processed as she still used the fast but imprecise translation technique of her first trial.

\textit{Final}. Stefania switches from the fast translation to the more precise one simply by configuration of the experimentation workbench. This technique uses a  different approach and produces different outputs, but the experimentation workbench handles format translations transparently. Hence, changing the complete analysis is again as easy as simply modifying a configuration option. This helps to avoid introducing accidental changes of the experiment that could occur if more complex programming would be involved. Stefania compares the final results with her pre-study. This is easy to do as she used the documentation feature, which results in automatic archiving of all input and output data, implementation artifacts, source code, and the entire configuration of the experimentation workbench. Apart from the impact of the more detailed analysis, the results match. Hence, Stefania shares the documentation file with her fellow researchers, who can directly rerun the analysis and compare the results or do further studies.

It is exactly this kind of fast, iterative changes along with the comprehensive documentation that the concept of an experimentation workbench is about.

\section{Concepts and Requirements}
\label{sec:requirements}
As illustrated in the usage scenario above, experimentation workbenches should support researchers in easily performing experiments, explore the space of possibilities, document them, and share them with others, who then can build on them, refine them, apply their own techniques or create further derived experiments. These goals partially overlap with other approaches to improve the scientific process in software engineering.

 For example, benchmarking as a scientific approach can support community building and can help to accelerate scientific advancement~\cite{SimEasterbrookHolt03}. However, it does not address aspects like replication, supporting the experimentation process itself, etc. Concepts like Jupyter notebooks~\cite{Jupyter} support experimentation and to some limited degree replication and sharing, so they already come close. We could consider them as one specific instance of an experimentation workbench for data science, but this is usually not applicable to software engineering experimentation and it is still very generic, leaving the major burden of programming to the researchers. Other concepts like using docker images or virtual machines in software engineering address replication~\cite{Boettiger15}, but not other experimentation-oriented capabilities. Thus,  while various approaches exist that address related topics, so far no one  fully addresses the problems of the software engineering researcher as we do here with the concept of experimentation workbenches. 

In our vision, experimentation workbenches provide key capabilities to support typical research activities in the scientific workflow. However, we do not expect that there will be a single experimentation workbench for all kinds of software engineering research just as there is no single experimentation facility in physics. Rather, we expect that the generic requirements, we present below, will be instantiated in domain-oriented ways. 
For clarity, we abstract here from any activities that are already well-supported, e.g., by development environments, and focus on those, for which  there is typically no automated support available. In our view these are, in particular, the following ones:

\begin{enumerate}[label=\RQLayerOne{\arabic*}]
  \item\label{req:setup} Support the setup (definition) of experiments. 
  \item\label{req:analysis} Support the analysis of experiments.
  \item\label{req:exp-variant} Support the fast execution of variants of the experiment, including applying the experiment setup to different cases.
  \item\label{req:doc} Support the documentation of all relevant artefacts for replication.
	\item\label{req:reuse} Support the reuse of experiments (by third parties).
	\item\label{req:extension} Support the extension and specialization of experiments by third parties. 
\end{enumerate}

Supporting the \textit{setup of experiments} (\ref{req:setup}) means, in particular, that technical issues that are not relevant to the study, but only required to ensure its execution, are handled by the workbench as far as possible. These could include providing initialization code, process coordination, and parallelization. Platform independence could be another aspect, which is not mandatory, but rather a design decision made by the developers and judged according to the requirements of the type of experimentation to be supported. Ideally, researchers only need to focus on the algorithmic aspects of their contributions. Thus, the front end to the researcher should provide a configuration interface or a Domain-Specific Language (DSL) or a combination of both to assist in these tasks. 

After an experiment execution an \textit{experiment analysis} (\ref{req:analysis}) must be done in order to determine what the results mean in relation to the initial research question. This could be provided by visualization tools, by providing certain kinds of tabularization, or simply by analysis scripts. The needs in this area are strongly domain-dependent as  the analysis will depend on the types and amounts of data produced, requirements on statistics, and so forth. 
However, in many cases it will be possible to address these requirements using environments for data analysis like R~\cite{R-Project}. Thus, if appropriate interfaces are available, there is no need to re-implement this for each workbench.

In experimentation it is often the case that one wants to \textit{analyze  variations in the data or in algorithms} to determine their impact on the overall outcome. This requires the possibility to set up new versions of an experiment with little effort and to easily go back to the previous analysis, if an experiment turns out to be not successful (\ref{req:exp-variant}). Sometimes such a variation can also be driven by performing a simplified version to improve turn-around time.

Finally, an experimentation workbench should support \textit{documentation of experiments} such that automated replication is easily facilitated (\ref{req:doc}). Such a replication package should at least include all inputs, outputs, code, and analysis results, if applicable. Thus, the package should directly support the inspection of any results, but also the direct replication of the experiments by any third-party. 

Ideally, it should be possible to directly \textit{reuse} not only the results, but even the experiments (\ref{req:reuse}). While this reusability enables repeatability by allowing researchers to always receive the same results with the same experimental setup, it also supports replicability and reproducibility by different persons. In particular, third parties should be able to easily re-conduct an experiment by reusing the experimental setup either directly, or with only slight adaptations, e.g., to fit their environment, which still conform to the initially documented experiment.

The direct reuse of an experiment (\ref{req:reuse}) may not always be sufficient to enable reproducibility. For example, if a third party aims at conducting a previous experiment of other researchers using a different case study. This may require variations like different algorithms to provide the necessary data from software artifacts as the new case study consists of different types of artifacts than the initial one (e.g., Java source code instead of C source code). This may require \textit{extensions or specializations} of the initial experimentation, which ideally should be directly supported by the experimentation workbench (\ref{req:extension}). Moreover, from the perspective of the overall scientific process that should be supported along the lines of the well-known adage of “standing on the shoulders of giants”, this requirement is actually particularly important. Today, such an extension is extremely difficult, even if all the code is available as open source as existing experimental implementations are typically not created for reuse or even extension by third-parties. Thus, we want to emphasize this here due its importance to the scientific process.

\section{An Experimentation Workbench for Static Product Line Analysis}\label{sec:implementation}

In this section, we  discuss KernelHaven\footnote{Available at GitHub: \url{https://github.com/KernelHaven/KernelHaven}} as an example of an open source experimentation workbench \cite{KroeherEl-SharkawySchmid18,KroeherEl-SharkawySchmid18b}.
We do not  argue that it is the ideal or perfect implementation of an experimentation workbench, but we use it here as a reference to describe some properties and technical implications of the concepts and requirements introduced in Section~\ref{sec:requirements}. KernelHaven instantiates these generic requirements for the domain of static analysis of software product lines. While we focus on this domain here, a specialized instance\footnote{Available at GitHub: \url{https://github.com/KernelHaven/MetricHaven}} of KernelHaven exists, which addresses metrics for software product lines~\cite{El-SharkawyYamagishi-EichlerSchmid19} as a subset of static product line analysis (cf. requirement~\ref{req:extension} in Section~\ref{sec:requirements}).

\begin{figure}[t] 
	\centering
		\includegraphics[width=0.8\columnwidth,trim={0cm 4,5cm 13,5cm 0cm},clip]{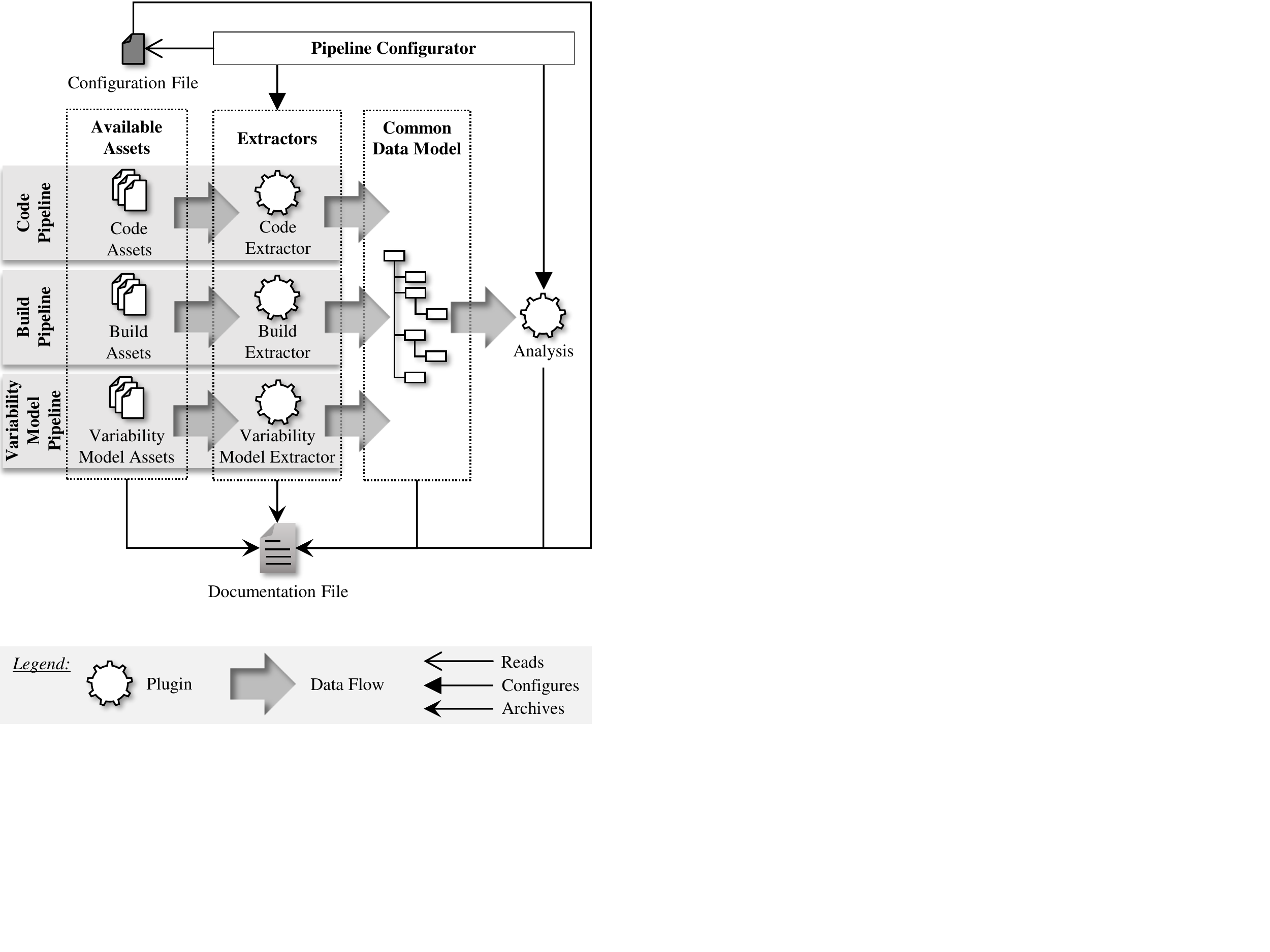}
	\caption{KernelHaven Overview.}
	\label{fig:Architecture}
\end{figure}

In order to abstract from technical details and allow to rapidly set up new experiment variants (\ref{req:exp-variant}), it is necessary to take a domain-oriented perspective. The resulting workbench will only support experiments in this domain. In our case of product line analysis, Figure~\ref{fig:Architecture} shows the resulting structure of that workbench. It consists of various \textit{extractors}, which transform the \textit{available assets} into a \textit{common data model} that provides a good basis for \textit{analysis}. In our domain, the relevant information is typically derived from three categories of assets: the variability model, the build system, and code assets. Hence, a \textit{code pipeline}, a \textit{build pipeline}, and a \textit{variability model pipeline} further structure the workbench in Figure~\ref{fig:Architecture}, which perform this derivation for the respective category of assets individually.

While this workbench was initially developed for experiments on Linux, its architecture is much broader as all analysis and extractor components are implemented by a flexible plugin system. Thus, for example,  the application to a proprietary variational build system only requires the development of an appropriate extraction plugin.\footnote{In the case of minor variations, of course, also variations of existing plugins or even parameterized instances can be used. In order to support this a parametrization approach for plugins exists.} The common data model, which is used to represent the collected data of the various extractors, allows the reuse of existing analysis plugins without additional  work. 

Figure~\ref{fig:Architecture} shows that the \textit{pipeline configurator} reads a \textit{configuration file} to configure the whole infrastructure. It performs initialization of all subsystems (in particular the wiring, initialization and starting of the components), creates the corresponding processes, and allocates hardware resources. Also issues like parallelization of the various processes are handled by the infrastructure. Initially an adequate number of processes are created and throughout data dependencies are used to manage the parallel processing. 

The configuration-oriented approach, which leads to an open ecosystem platform, directly addresses requirements~\ref{req:setup} to~\ref{req:exp-variant} (cf. Section~\ref{sec:requirements}). The platform can also be configured to directly invoke documentation-related activities like archiving all relevant data, sources, implementations, configuration information, and so forth. This addresses requirement~\ref{req:doc} in Section~\ref{sec:requirements}. Requirements~\ref{req:reuse} and~\ref{req:extension} are  addressed by combining that (a) other researchers can rerun the experiments due to auto-documentation and (b) build on them by changing the configuration using either existing or self-developed plugins. The auto-documentation feature of KernelHaven therefore produces the experiment documentation in terms of an archive that contains all input, intermediate, and output data, as well as the main infrastructure, all plug-ins, and the configuration file. This feature directly supports reproducibility as a core task of research: the archive provides the original experimental setup to other researchers, enables them to rerun the same experiment on the same input data, and allows inspection of the previous results.

Initially, KernelHaven plugins were mostly derived from existing research prototypes. For example, they wrap a pre-existing tool and handle all the details of driving these tools (e.g., particular parametrization or environment needs). This has two major effects:
\begin{itemize}
	\item The (re-)use of successful tools has been tremendously simplified: while for some tools, like TypeChef~\cite{typechef}, people typically need several days to make it work reliably, the plugin embeds the relevant knowledge to make it reusable in minutes.
	\item The combination of tools is now possible simply by configuration: while combining existing tools as well as integrating with existing ones requires a lot of work and tool-knowledge, it is now a matter of defining the desired plugins as a parameter in a configuration file.
\end{itemize}  

An important part of the domain design is the definition of the data structures and relevant data transformations to make extractors interchangeable. This can also be illustrated with  KernelHaven. The toolset provides several extractor plugins, which can operate on C-Source code and can provide variability-tagged source-code fragments. One is derived from Undertaker~\cite{Undertaker}, another one from TypeChef~\cite{typechef}. They differ, however, very significantly in terms of the level of detail they provide: Undertaker scans the source-file, identifies code blocks as sequences of lines and tags them with the relevant variability derived from any \texttt{\#ifdef}-command. In the process it ignores header-files. On the other hand, TypeChef performs full variability-aware parsing, including header-files and macro expansions. As a consequence, it provides a complete AST adorned with variability information.

While the results of the two tools differ fundamentally, they share some information. Both extract the included variability information from source-files using preprocessor directives, i.e., the \textit{presence conditions}.
This commonality is sufficient for some types of analyses, like the identification of dead code \cite{TartlerLohmannSincero+11}. In KernelHaven, all entities for representing extracted code information inherit from a class, which stores this common information. This allows to exchange code extractors as long as the desired analysis does not require the specific output of a certain extractor. The plugin system knows about these dependencies and takes care of them. An example is illustrated in Listing~\ref{lst:Configured-Pipeline}, which shows the relevant part to perform a dead code analysis on Linux with the Undertaker-extractor. Only the configuration file, in particular Line~1, must be modified in order to use TypeChef instead of Undertaker. 
However, this can be seen as a refinement of the Undertaker-information as this also corresponds to code-blocks. This is actually how the information is  represented: a source-code processor may provide variability-adorned code-blocks, which may contain more detailed information (e.g., AST). The data structures are defined in a way that further steps may ignore levels of detail that are not required in their processing increasing composability of the various plugins. 

While the analysis of results itself is not part of KernelHaven, the infrastructure supports \ref{req:analysis} by supporting the export of the resulting data in analysis-friendly formats like text-files (e.g., csv) or Excel.  The core analysis is then typically done either with Excel or using R-scripts.

This allows to execute the scenario described in Section~\ref{sec:scenario}. One can first test new analysis concepts based on the rather fast, but not so detailed Undertaker-extractor, which extracts variability elements as line ranges. After the analysis has been positively evaluated, one can perform a more detailed analysis using the macro-aware parser of TypeChef, which provides a code block as an AST-fragment where all elements have the same presence condition. So, what both extractors have in common is to provide source code elements tagged with presence conditions, which is sufficient for dead code analysis. An AST-based analysis like type-analysis requires code extractors, which extract an AST containing variability information. Currently KernelHaven supports this with TypeChef or srcML~\cite{srcML}. It is important to note that (a) these different types of analysis are all supported by KernelHaven, and (b), for switching among them, it is sufficient to change some configuration options; no implementation change for any extractor plugin or the analysis plugin is required as long as they all adhere to the interface conventions.

\begin{figure}[tb]
	\centering
\begin{lstlisting}[language=properties,label=lst:Configured-Pipeline,caption={Excerpt of a KernelHaven configuration file.}]
code.extractor.class =	UndertakerExtractor
code.extractor.file_regex =	.*\.c
build.extractor.class =	KbuildMinerExtractor
variability.extractor.class =	KconfigReaderExtractor
analysis.class =	DeadCodeAnalysis
...
\end{lstlisting}
\end{figure}

Here, KernelHaven realizes two different perspectives on experimentation workbenches. On the one hand, KernelHaven is a platform that provides support for various experiments in the domain of static product lines analyses. On the other, we derived different KernelHaven instances based on this platform. These instances consist of the common experimentation workbench, configuration parameters, and if necessary also experimentation-specific plugins that realize one specific analysis. In Section~\ref{sec:experiences}, we exemplary show some of the experiments, i.e., KernelHaven instances, which we realized based on the KernelHaven platform. 

\section{Challenges}\label{sec:discussion}

While we believe the concept of experimentation workbenches is very fruitful for the research community and our own experiences with the KernelHaven implementation  of it are so far very positive, there are still some challenges associated with the realization of an experimentation workbench. 

\textit{Domain specifity.} The first and most obvious challenge to experimentation workbenches is that they need to be constructed for a specific domain of experimentation. Thus, the requirements, we presented here, must be interpreted in the corresponding context and the capabilities of the workbench need to be scoped in terms of types of experiments (variations) to take into account. Thus, an experimentation workbench can be regarded as some form of product line~\cite{LindenSchmidRommes07} or open ecosystem~\cite{Bosch09}.
Similar to product lines, of course, incremental development of it is possible. 

\textit{Freedom of Implementation.}
It is fundamentally hard to guarantee full replicability, without significantly restricting the expressiveness used for realising specific parts of an experimental implementation. This is particularly the case for an experimentation workbench, like KernelHaven, that even allows pre-existing systems written in different languages and with arbitrary infrastructures as plugins. This issue is further compounded as in different domains different aspects may be important for replicability. For example, KernelHaven is purely functionality-related, i.e., as long as the same outputs are achieved for the same input, we can assume replicability. In other areas like performance engineering, the issue is different as similar timing behavior is required for replicability~\cite{EichelbergerSassSchmid16}. Hence, a corresponding experimentation workbench will have to address different issues. In this special case, special performance-rated environments have been proposed to promote replication~\cite{OliveiraPetkovichReidemeister+13}. 

\textit{Scope of Documentation.}
An important issue is the scope of an  implementation that needs to be archived  for replicability. 
In the example given in Section~\ref{sec:implementation} only code artifacts related to the workbench implementation and the plugins are considered. The Java virtual machine and the operating system are not included. This yields rather lightweight packages where multiple archives can  easily be stored locally. However, in other contexts the replication may require a copy of the virtual machine and the operating system. 
In such cases, an experimentation  workbench may of course directly create a docker image or a virtual machine~\cite{Boettiger15}.
One can even imagine cases where a complex multi-machine setup needs to be archived like in large-scale adaptive systems.


\textit{Controlled Experiment Variation.}
A related issue are experiment variations. If some part of the analysis is replaced by something else, then this will result in changes to the experiment. Typically,   this will also invoke undesirable changes. In the example given, a switch from the simple code extractor to the more detailed one does not only lead to a more precise analysis of variability information in header files, but can also impact the details of the analyzed blocks as they are analyzed in a different way. Whether these changes are acceptable or not, will depend very much on the specifics of the analysis performed. In case plugins are used that have been engineered from the beginning with an experimentation workbench in mind, we expect this also to be less of an issue than it is currently the case with the reengineered plugins that KernelHaven uses.

These challenges basically come down to the need of achieving a sufficient domain understanding. Either prior to the construction of such an environment or as part of the experimental process. In this regard the development of an experimentation workbench can be compared to the development of a software product line.
\section{Experiences}\label{sec:experiences}
So far, we described the general requirements for experimentation workbenches and how the research community can take advantage of them. In this section, we share our experiences when working with KernelHaven (cf.\ Section~\ref{sec:implementation}). We used this experimentation workbench for our own research in the ITEA3 project REVaMP2\footnote{\url{http://www.revamp2-project.eu/}}, which focuses on round-trip engineering of software product lines. Since KernelHaven supports the definition of various experiments (\ref{req:setup}) in the domain of static SPL analysis, we were able to use KernelHaven for many different research activities, like for example reverse engineering of variability information for bootstrapping of SPL development, evolution support, and verification tasks. We provide an overview of the variety of analyses supported by KernelHaven and show how we could realize these very diverse analyses with limited development resources in a short time. Further, we present lessons learned when working with KernelHaven.

Together with the Robert Bosch GmbH, we worked on reverse engineering of a dependency management system for a large-scale industrial product line \cite{El-SharkawyDharKrafczyk+18}. For this, we decided to adapt the feature effect analysis described by  Nadi et al.\ \cite{NadiBergerKastner+15} to the needs of Bosch. This kind of analysis requires usually much effort to combine various parsers that extract variability information from different information sources. By means of KernelHaven, we were able to develop a first prototype very quickly, since the  combination of data from different sources is a major concern of KernelHaven and first suitable parsers were already present. As a result, we could focus on the integration of parsers specific to the development environment of Bosch \cite{El-SharkawyDharKrafczyk+18}, lifting the propositional analysis of feature effects to integer-based variability \cite{KrafczykEl-SharkawySchmid18}, and on providing visualization support for reverse engineered dependencies \cite{GruenerBurgerAbukwaik+19}. In addition, KernelHaven's reproduction support (\ref{req:doc}) simplified the execution of configured algorithms at the two partners. Thus, we also achieved a significant benefit for industrial transfer of our research results.

KernelHaven also supports  the verification of various  properties of SPLs through its data extraction and analysis capabilities. For instance, we can reproduce and freely combine a large number of published product line metrics \cite{El-SharkawyKrafczykSchmid19} resulting in more than 23,000 variations of metrics for single systems and SPLs, many of which are not handled by any other tool \cite{El-SharkawyYamagishi-EichlerSchmid19}. Another very important aspect for SPLs, is the analysis of (un-)dead code with respect to its variability model \cite{TartlerLohmannSincero+11}. This is a very time consuming task as it analyzes whether the variability model allows the (de-)selection of all configurable code parts, e.g., \texttt{\#ifdef}-blocks. Thus, this kind of analysis is more suitable for daily builds than for a continuous analysis during the development. However, a commit analysis of the Linux kernel has shown that changes to variability information occur infrequently and only affect small parts \cite{KroherGerlingSchmid18}. Based on this insight, we implemented an incremental verification approach to reduce the overall time consumption by about 90\% \cite{Floeter18}, which is suitable to be applied in a continuous development environment. The incremental verification is realized by combining previous results of an already available dead code analysis with a new analysis that detects changed variability information (\ref{req:reuse} and~\ref{req:extension}).

Through the broad range of conducted experiments in combination with tested variations of algorithms, KernelHaven evolved quickly to a highly configurable system. For this, we realized a documentation system that provides the user, based on installed plugins, a list of available configuration options, supported values, and default settings. However, this system does not scale well as it neither supports a documentation of suggested settings arising through the combination of multiple plugins nor does it provide a dependency management among the plugins, e.g., the metric analysis plugin requires code extractors that extract a variability-aware AST rather than a simple block structure as needed by most other analyses (cf. Section~\ref{sec:implementation}). Thus, for the future, we plan to address this issue by 1.) limiting the amount of configuration possibilities for stable plugins and by 2.) integrating dependency management systems suitable for software ecosystems~\cite{BrummermannKeuneckeSchmid12a}, e.g., based on our EASyProducer implementation~\cite{SchmidEichelberger15a}\footnote{https://sse.uni-hildesheim.de/en/research/projects/easy-producer/}.

This does also  strongly suggest that experimentation workbenches can be regarded as a special form of product line or open software ecosystem \cite{Schmid10a}. 

\section{Conclusion}\label{sec:conclusion}

In this paper, we introduced the concept of an experimentation workbench as a way of thinking about scientific experimentation artifacts with a focus on the needs of the scientific process. We believe that thinking about experimental research software in terms of this concept provides significant advantages when developing research systems in software engineering. In the future, we believe that some powerful experimentation workbenches for specific software engineering domains may provide a major contribution and foster the development of better ecosystems that drive software engineering research. 

Our main contributions besides the concept itself are the characterizing requirements, which define an ``ideal'' experimentation workbench along with an illustrative scenario. We further described KernelHaven as an example experimentation workbench situated in the domain of product line analysis. KernelHaven may provide a basis for a research ecosystem for product line analysis as  it integrates already today a number of existing research tools and makes them significantly more accessible than is otherwise the case. 
Besides achieving already significant research benefits, as discussed, we also found that this approach significantly improves our potential of working with industrial partners.

We assume that the concept of an experimentation workbench always needs to be interpreted relative to the specific scientific area. However, we hope the general requirements we presented may guide the creation of such systems and thus support the scientific progress by fostering the creation of ecosystems around experimentation workbenches in a number of software engineering fields. For example, one may interpret our concept presented in this paper in the context of Natural Language Processing (NLP) in requirements engineering~\cite{GnesiTrentanni19,GnesiFerrari18,FerrariDell-OrlettaEsuli+17}. In particular, the NLP tool for requirements analysis~\cite{GnesiTrentanni19} may provide an excellent foundation for extending it to an experimentation workbench for that domain in future.

\section*{Acknowledgements}
This work is partially supported by the ITEA3 project REVaMP$^2$, funded by the BMBF (German Ministry of Research and Education) under grant 01IS16042H. Any opinions expressed herein are solely by the authors and not by the BMBF.

\bibliographystyle{spmpsci}
\bibliography{literature}

\end{document}